\UseRawInputEncoding
\UseRawInputEncoding
\documentclass[%
 reprint,
 superscriptaddress,
 amsmath,amssymb,
 aps,
prl 
]{revtex4-2}

\usepackage[pagewise,modulo]{lineno}
\usepackage{xcolor}
\usepackage{graphicx}
\usepackage{dcolumn}
\usepackage{bm}
\usepackage{chemformula} 
\usepackage[hidelinks,colorlinks=true,linkcolor=blue,urlcolor=blue,citecolor=blue]{hyperref}

\makeatletter
\newcommand{\mycaption}[2][]{%
  \begingroup%
  \renewcommand{\figurename}{\textbf{Figure}}
  \renewcommand{\@caption@fignum@sep}{ \textbf{$\vert$} }%
  \renewcommand{\fnum@figure}{{\normalfont\bfseries \figurename~\thefigure}}
  \caption[#1]{#2}%
  \endgroup%
}
\makeatother

\makeatletter
\def\ignorecitefornumbering#1{%
     \begingroup
         \@fileswfalse
         #1
    \endgroup
}
\makeatother

\begin{document}

%
%
%
%
%
%
%
%
%
%


\title{Proximity Magnetism in \ch{Mn(Bi,Sb)2Te4}--\ch{(Bi,Sb)2Te3}/MnTe Natural Heterostructures}
\author{Owen A. Vail}    
    \email{owen.a.vail.civ@army.mil}
    \affiliation{DEVCOM Army Research Laboratory, Adelphi, Maryland 20783, USA}
\author{Shu-Wei Wang}    
    \affiliation{Francis Bitter Magnet Laboratory, Plasma Science and Fusion Center, Massachusetts Institute of Technology, Cambridge, Massachusetts 02139, USA}
    \affiliation{Department of Electrical Engineering, National Cheng Kung University, Tainan 70101, Taiwan}
\author{Yasen Hou}    
    \affiliation{Francis Bitter Magnet Laboratory, Plasma Science and Fusion Center, Massachusetts Institute of Technology, Cambridge, Massachusetts 02139, USA}
\author{Dinura Hettiarachchi}    
    \affiliation{Department of Physics, University of Ottawa, Ottawa, Ontario K1N 6N5, Canada}
\author{Jean-F\'{e}lix Milette}
    \affiliation{Department of Physics, University of Ottawa, Ottawa, Ontario K1N 6N5, Canada}
\author{Tim B. Eldred}    
    \affiliation{Department of Materials Science and Engineering, North Carolina State University, Raleigh, North Carolina 27695, USA}
\author{Wenpei Gao}    
    \affiliation{Department of Materials Science and Engineering, North Carolina State University, Raleigh, North Carolina 27695, USA}
\author{Wendy L. Sarney}
    \affiliation{DEVCOM Army Research Laboratory, Adelphi, Maryland 20783, USA}
\author{Haile Ambaye}    
    \affiliation{Neutron Scattering Division, Neutron Sciences Directorate, Oak Ridge National Laboratory, Oak Ridge, Tennessee 37831, USA}
\author{Jong Keum}    
    \affiliation{Neutron Scattering Division, Neutron Sciences Directorate, Oak Ridge National Laboratory, Oak Ridge, Tennessee 37831, USA}
    \affiliation{Center for Nanophase Materials Sciences, Physical Science Directorate, Oak Ridge National Laboratory, Oak Ridge, Tennessee 37831, USA}    
\author{Valeria Lauter}
    \email{lauterv@ornl.gov}
    \affiliation{Neutron Scattering Division, Neutron Sciences Directorate, Oak Ridge National Laboratory, Oak Ridge, Tennessee 37831, USA}
\author{George J. de Coster}
    \affiliation{DEVCOM Army Research Laboratory, Adelphi, Maryland 20783, USA}
    \affiliation{Institute of Physics, Faculty of Electrical Engineering and Information Technology, University of the Bundeswehr Munich, Neubiberg 85577, Germany}
\author{Matthew J. Gilbert}
    \affiliation{Department of Electrical and Computer Engineering, University of Illinois at Urbana-Champaign, Urbana, Illinois 61801, USA}
\author{Don Heiman}    
    \affiliation{Francis Bitter Magnet Laboratory, Plasma Science and Fusion Center, Massachusetts Institute of Technology, Cambridge, Massachusetts 02139, USA}
    \affiliation{Department of Physics, Northeastern University, Boston, Massachusetts 02115, USA}
\author{Jagadeesh S. Moodera}    
    \email{moodera@mit.edu}
    \affiliation{Francis Bitter Magnet Laboratory, Plasma Science and Fusion Center, Massachusetts Institute of Technology, Cambridge, Massachusetts 02139, USA}
    \affiliation{Department of Physics, Massachusetts Institute of Technology, Cambridge, Massachusetts 02139, USA}
\author{Hang Chi}    
    \email{hang.chi@uottawa.ca}
    \affiliation{Department of Physics, University of Ottawa, Ottawa, Ontario K1N 6N5, Canada}
    \affiliation{School of Electrical Engineering and Computer Science, University of Ottawa, Ottawa, Ontario K1N 6N5, Canada}
    \affiliation{Nexus for Quantum Technologies, University of Ottawa, Ottawa, Ontario K1N 6N5, Canada}

\date{\today}

\begin{abstract}
Magnetic topological insulators and their heterostructures provide great opportunities in coupling band topology with nontrivial spin configuration for enhanced spintronic device performance as well as designing totally new magnetoelectric systems and functionalities. We find that Mn interdiffusion from MnTe when interfaced with \ch{(Bi,Sb)2Te3} stabilizes as self-organized \ch{Mn(Bi,Sb)2Te4} septuple lamellae amongst alternating \ch{(Bi,Sb)2Te3} quintuple layers, as observed using scanning transmission electron microscopy and depth-sensitive polarized neutron reflectometry. We further demonstrate a valuable combination of magnetic and topological orders in these naturally formed \ch{Mn(Bi,Sb)2Te4}--\ch{(Bi,Sb)2Te3} heterostructures that are exchange coupled with MnTe. Magnetotransport experiments and quantum magnetism simulations reveal that, above its own N\'eel temperature $T_{\rm N} \sim$ 20 K, \ch{Mn(Bi,Sb)2Te4} mediates the exchange field leading to an anomalous Hall effect at the \ch{(Bi,Sb)2Te3}/MnTe interface, with an enhanced interfacial $T_{\rm N}$ exceeding~200~K, approaching that of the bulk MnTe. This novel magnetic interface in turn allows a robust and deterministic spin-orbit torque switching without an external magnetic field at a low critical current density of $3\times10^{5}$~A~cm$^{-2}$. The antiferromagnetically coupled architecture of \ch{Mn(Bi,Sb)2Te4}--\ch{(Bi,Sb)2Te3}/MnTe, featuring unique magnetic and topological proximity effects across a chalcogenide backbone, is rich in fundamental interface physics and holds potential for practical applications in spintronics.
\end{abstract}

\maketitle

\begin{figure*}
\includegraphics[width=1\textwidth]{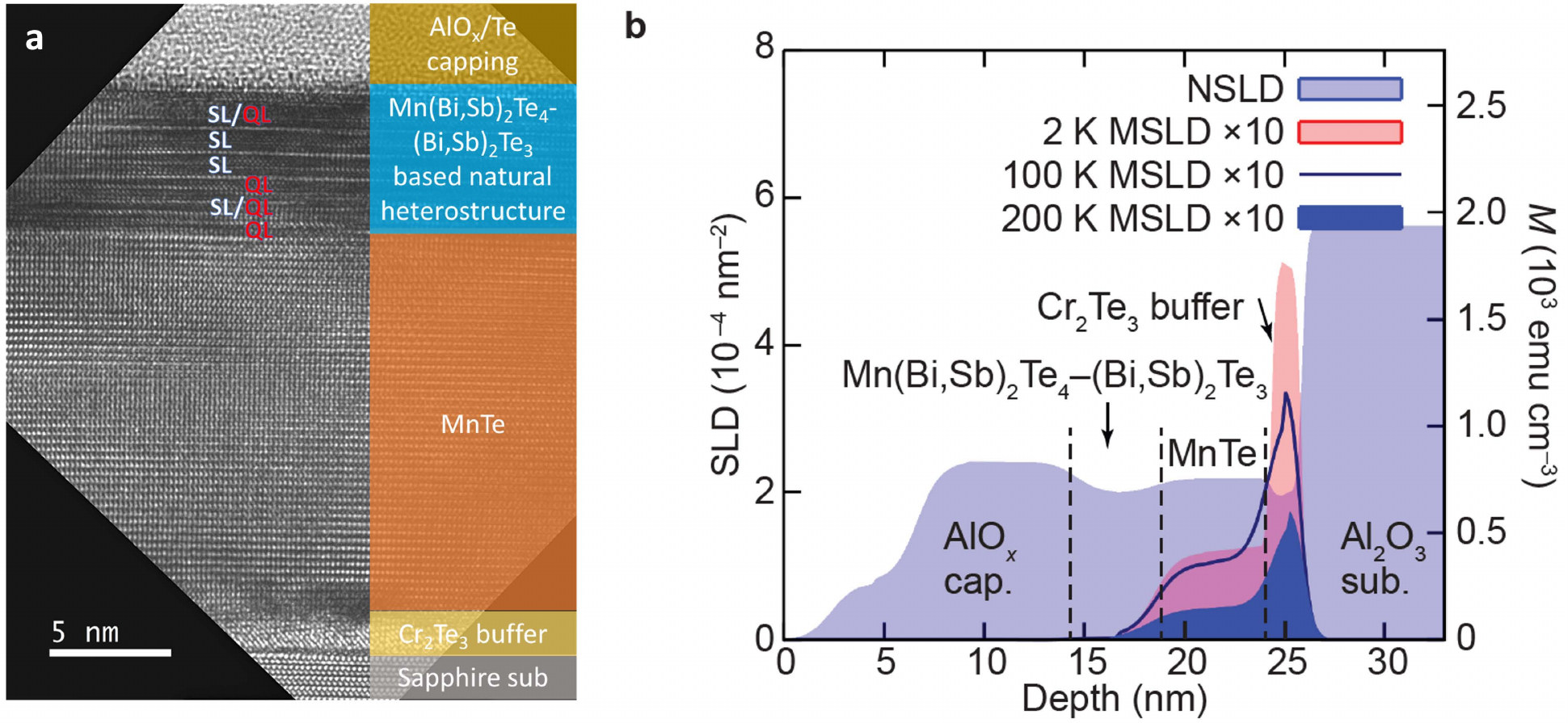}
\mycaption{\label{fig:fig1}\textbf{Crystal and magnetic structure of \ch{Mn(Bi,Sb)2Te4}--\ch{(Bi,Sb)2Te3}/MnTe natural heterostructure.} \textbf{a} Cross-sectional STEM image of the natural heterostructure grown on \ch{Al2O3}(0001) substrate with \ch{Cr2Te3} buffer layer, where \ch{Mn(Bi,Sb)2Te4} septuple layers (SLs) are dispersed in the matrix of \ch{(Bi,Sb)2Te3} quintuple layers (QLs). \textbf{b} Depth profiles of PNR nuclear (NSLD) and magnetic (MSLD) scattering length densities (SLD) at selected temperature of 2, 100 and 200~K, under an in-plane magnetic field $\mu_0H$ = 4.5 T.}
\end{figure*}

\noindent\textbf{Introduction} 

The union of topology and magnetism in condensed matter systems provides intriguing opportunities for unprecedented electrical control of dynamic orders \cite{Bernevig2022}. Spin-orbit interaction in particular allows for symmetry-protected chiral spin-textures in topologically nontrivial bands that can be leveraged for dissipationless quantum transport \cite{Chang2023}. The persistence of magnetic order is well complemented by efficient spin manipulation, such as that found in spin-orbit coupled topological insulator (TI) \ch{(Bi,Sb)2Te3} (BAT) \cite{Heremans2017}. Rationally designed magnetoelectric platforms offer a robust framework for effective information storage and spintronic manipulation, vital for highly efficient and deterministic spin-orbit torque (SOT) applications \cite{Matsukura2015,Manchon2019,He2022}.

Synergizing magnetic and topological functionalities across interfaces allows for designer heterostructures with salient properties including gap size, magnetic resonance frequency, and critical temperature \cite{Chi2022}. While doping magnetic elements into a TI has been successful in opening a surface exchange gap \cite{Chang2013}, challenges remain for creating long-range order at higher temperatures without deteriorating the material quality. \ch{MnBi2Te4}-based materials represent a promising new family for exploring magnetic topological phenomena, featuring nontrivial topology and out-of-plane (OOP) interlayer antiferromagnetic (AF) ordering with a N\'eel temperature $T_{\rm N}$ $\sim$ 20~K. The quantum anomalous Hall effect has been recently observed, offering viable pathways elevating the full quantization temperature beyond the current sub-Kelvin range \cite{Otrokov2019, Rienks2019, Deng2020, Deng2021}. Furthermore, the in-plane (IP) antiferromagnetic/altermagnetic order in MnTe \cite{Kriegner2016NC_Multiplestable, Yin2019PRL_Planar, Krempasky2024} and its effective coupling to BAT layers bode well for substantial further tuning of transport behavior towards practical applications. 

Here we demonstrate that magnetic topological insulator (MTI) \ch{Mn(Bi,Sb)2Te4} (MBAT) septuple layers naturally self-assemble at BAT/MnTe interface during molecular beam epitaxy (MBE) growth, constituting an exchange-coupled topological stack that transcends the constituent materials \cite{Katmis2016}. Magneto-transport and neutron data identify synergistic magnetic and spin-orbit orders, consistent with a BAT/MnTe interface featuring OOP magnetization stabilized by non-local MTI, as revealed by quantum magnetism simulations. Robust spin torque responses in absence of external magnetic field prove viable for customizable SOT devices with low switching current density and energy consumption.

\begin{figure*}
\includegraphics[width=1\textwidth]{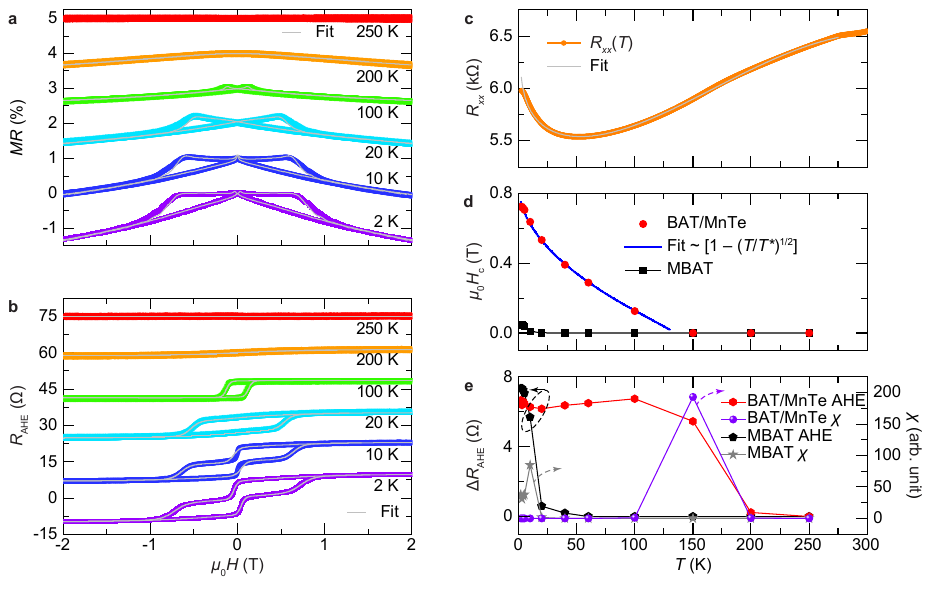}
\mycaption{\label{fig:fig2}\textbf{Magneto-transport properties of \ch{Mn(Bi,Sb)2Te4}--\ch{(Bi,Sb)2Te3}/MnTe.} \textbf{a}-\textbf{b} Magnetic field dependence of the magnetoresistance $MR \equiv [R_{xx}(H)-R_{xx}(0)]/R_{xx}(0) \times 100\%$ (\textbf{a}) and the anomalous Hall resistance $R_{\rm AHE}$ (i.e. the Hall resistance $R_{yx}(H)$ after removing the linear-$H$ background) at selected temperatures in the range of $T$ = 2 -- 250 K. The grey lines in \textbf{a} and \textbf{b} illustrate the numerical fits using Eq. \eqref{eq2} for convoluted magnetic hysteresis and weak antilocalization and Eq. \eqref{eq3} for multiple hysteretic loops, respectively. For clarity, the curves are vertically shifted by 1\% (15 $\Omega$) in \textbf{a} (\textbf{b}). \textbf{c} Temperature dependent longitudinal resistance $R_{xx}(T)$ with a minimum around 50 K, well characterized by the numerical fit using Eq. \eqref{eq1}. \textbf{d}-\textbf{e} Temperature dependence of the coercive fields (\textbf{d} symbols), the magnitude of the anomalous Hall effect $\Delta R_{\rm AHE}$ (\textbf{e} left axis) and the susceptibility $\chi$ derived from Eq. \eqref{eq3} (\textbf{e} right axis), revealing competing magnetic components, where the BAT/MnTe interface displays a critical behavior with an exponent of $\mathrm{1/2}$  (\textbf{d} line).}
\end{figure*}

\noindent\textbf{Crystal and Magnetic Structure}

Figure~\ref{fig:fig1}a illustrates a cross-sectional view of the micro structure of the MBAT-BAT/MnTe heterostructure with an optimal \ch{Cr2Te3} buffer on \ch{Al2O3}(0001) substrate. It is evident that the interface between MnTe and BAT layers is susceptible to Mn diffusion (Fig. S1), via e.g. grain boundaries and across the van der Waals gaps, as often seen in magnetic TI heterostructures \cite{Awana2022, Bhattacharjee2022AM_Topological}. The structural non-uniformity under the present growth conditions results in naturally intermixed MBAT-BAT layers (Fig. S2). Bulk $\alpha$-MnTe (IP) and \ch{MnBi2Te4} (OOP) are antiferromagnets (AFMs) with $T_{\rm N}$ of 310 K \cite{Li2022} and 25~K \cite{Otrokov2019}, respectively. Earlier work on BAT/MnTe bilayers has shown anomalous Hall effect (AHE) below 240~K, displaying proximitized OOP magnetic moments \cite{He2018}, but without the important exchange-coupled MBAT layers. 

We have performed temperature and magnetic field dependent depth-sensitive polarized neutron reflectometry (PNR) experiments to uncover the magnetic structure and anisotropy of these films. As shown in Fig.~\ref{fig:fig1}b, the depth profile of the nuclear scattering length density (NSLD) is overall uniform, suggesting the high quality of the film. The magnetic scattering length density (MSLD) profiles were collected under IP magnetic field $\mu_{0}H$ = 4.5~T above the spin-flop transition of MnTe, at seleted temperature $T = $ 2 K, 100 K and 200~K. Robust magnetism has been confirmed, where it is evident the MBAT-BAT stack displays interface-driven exchange coupling at technologically relevant temperatures exceeding 200~K. Importantly, the choice of an ultrathin ferromagnetic \ch{Cr2Te3} buffer promotes the preferential alignment of the magnetic moments in the antiferromagnetic MnTe. The PNR method has also been used to measure the perpendicular magnetic anisotropy (PMA), which is based on comparing the behavior of the sample at high IP magnetic fields (when the magnetic moments are forced to orient themselves within the plane) and at low IP magnetic fields (when the magnetic moments are oriented perpendicular to the plane) \cite{Bai2021APL_Polarized,Zhu2012APL_Study,Zhu2016JPCS_Study,Kirichuk2025_Origin,Ji2011PRB_Perpendicular}. Experiment in a very small magnetic field allows to measure possible component of the magnetization vector in the plane of the sample, or to prove the absence of such a component (no spin-flip scattering). If spin-flip scattering is present it contributes to the reflectivity \cite{Blundell1995PRB_Spinorientation,Fields2026AAMI_NonAltermagnetic} and/or off-specular scattering \cite{Toperverg2001PBCM_3D, pnrLauter2012PSACR_215}. 
Field-dependent PNR measurements (Fig. S3) have confirmed PMA in our sample. PNR experiments on an additional sample (Fig. S4) further attests to the reproducibility of our observation. The novel proximity-driven multilayer structure enables the engineering of rich magneto-electric responses.

\noindent\textbf{Transport Properties} 

MBAT-BAT/MnTe heterostructures were fabricated into 10 $\mu$m~$\times$~30 $\mu$m Hall bar devices and measured in a four-point configuration down to 2 K. As shown in Fig.~\ref{fig:fig2}c, upon cooling, the longitudinal electrical resistance $R_{xx}(T)$ decreases metallically, through a kink associated with the N\'eel transition of MnTe near 275~K. A Kondo minimum in $R_{xx}(T)$ develops at $\sim$~50~K, due to the interaction between charge carriers and magnetic moments in MnTe and MBAT \cite{Kondo1964}. The magnetic and Mn $d$-orbital effects are well captured by a modified Bloch--Gr\"uneisen--Mott (BGM) model \cite{Mott1958}, 
\begin{equation}
R_{xx}(T) = R_0 + \alpha T^2 - \beta T^3 + \delta \ln(1/T),\label{eq1}
\end{equation}
where $R_0$ = $6.3\times10^3$ $\Omega$ is the residual resistance, $\alpha$ = $4.5\times10^{-2}$~$\Omega {\rm K}^{-2}$ refers to the electron-electron interaction, $\beta$ = $9.7\times10^{-5}$~$\Omega{\rm K}^{-3}$ is the Mott contribution responsible for the inflection around 150~K, and $\delta$ = $2.2\times10^2$~$\Omega$ depicts the Kondo effect at low temperature, consistent with the scaling of weak localization in a TI \cite{Lu2014PRL}. This fitting is overall consistent with a $p$-type bulk-conducting TI with spin-orbit coupling and low temperature magnetic ordering.

As illustrated in Fig.~\ref{fig:fig2}a-b, the development of magnetism in the MBAT-BAT/MnTe hybrid is further corroborated by (i) the butterfly-shaped $R_{xx}(H)$ profiles and (ii) the hysteresis loops in the Hall resistance $R_{yx}(H)$ under an OOP field. Specifically, the negative magnetoresistance $MR \equiv [R_{xx}(H)-R_{xx}(0)]/R_{xx}(0)$] is well characterized by the Hikami--Larkin--Nagaoka (HLN) model \cite{Hikami1980} with $\alpha=1/2$ corresponding to weak localization,
\begin{eqnarray}\label{eq2}
\Delta G_{xx}(B) = \frac{\alpha e^2}{2\pi^2 \hbar}\left [\psi{\left(\frac{1}{2}+\frac{\hbar}{4  e |B| \mathcal{L}_i^2}\right)}-\ln \left(\frac{\hbar}{4  e |B| \mathcal{L}_i^2}\right)\right].\notag\\ 
\end{eqnarray}
Here $\Delta G_{xx}(B) = G_{xx}(B) - G_{xx}(0)$ is the relative change in the longitudinal conductance $G_{xx} = R_{xx}/(R_{xx}^2+R_{yx}^2)$, magnetic induction $B = \mu_0 (H+M)$ with the functional form of magnetization $M$ extracted from $R_{yx}(H) \propto M(H)$, $e$ is the elementary charge, $\hbar$ is the reduced Planck's constant, $\psi$ is the digamma function and $\mathcal{L}_i$ is the phase coherence length \cite{Stephen2020}. At 2 K a high-field fit of Eq. \eqref{eq2} to the data in Fig. \ref{fig:fig2}a yields $\mathcal{L} \sim$ 26 nm (see Fig. S5 for its temperature dependence). 

It is clear in Fig.~\ref{fig:fig2}b that, after removing the linear-$H$ background in $R_{yx}(H)$, the hysteretic anomalous Hall resistance $R_{\rm AHE}(H)$ is consistent with a two-component AHE originating from MBAT and BAT/MnTe. The overall field dependence is well described by
\begin{eqnarray}\label{eq3}
    R_{\rm AHE}(H) = 
    \frac{\mu_0}{p e}\sum_{i=1,2}  M_{{\rm s},i} \tanh\left[(H-H_{{\rm c},i})/H_{\rm m}\right],
\end{eqnarray}
where $p = 2.3\times10^{20} \text{~cm}^{-3}$ is the carrier density as determined from the linear-in-$H$ ordinary Hall effect dominant at high magnetic field up to 18 T, which is temperature insensitive in the range of 2 K to 300~K, $M_{\rm s}$ is the saturation magnetization, $H_{\rm c}$ is the coercive field (Fig.~\ref{fig:fig2}d, following a $1-(T/T^*)^{1/2}$ critical exponent behavior for the BAT/MnTe interface) and $H_{\rm m}$ is a normalization factor. The remanent anomalous Hall resistance $\Delta R_{\rm AHE}$ for MBAT (black pentagon) and BAT/MnTe (red hexagon) are extracted in the left axis of Fig.~\ref{fig:fig2}e, indicating the magnetic ordering emerges in these components below $T_{\rm N}$ $\sim$ 20 K and $\sim$ 200 K, respectively. Furthermore, as depicted in the right axis of Fig.~\ref{fig:fig2}e, the differential susceptibility of magnetization, $\chi = \partial_H M(H)|_{H=0}$ is found to peak at a characteristic critical temperature $T^*$  ($< T_{\rm N}$) of $\sim$ 10~K and $\sim$ 150~K, for MBAT (grey star) and the BAT/MnTe (purple circle) interface, respectively. These results are repeatable and consistent with IP magnetotransport (Figs. S6 and S7).

\begin{figure}
\includegraphics{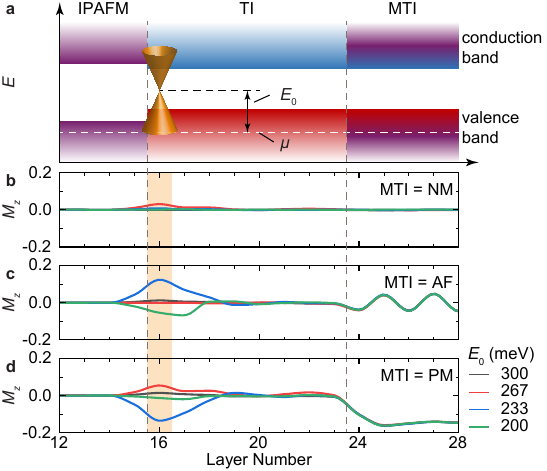}
\mycaption{\label{fig:fig3}\textbf{Quantum Magnetism Simulations.} \textbf{a} Band schematics of the generic model systems featuring three blocks consisting of 15 layers of in-plane ordered antiferromagnet (IPAFM, left), 8 layers of topological insulator (TI, middle), and $10$ layers of magnetic topological insulator (MTI, right). \textbf{b}-\textbf{d} Simulated out-of-plane layer averaged magnetization $M_z$ with the MTI block configured as non-magnetic (NM, \textbf{b}), antiferromagnetic (AF, \textbf{c}), and paramagnetic (PM, \textbf{d}). The relative band alignment $E_0$ of the TI block varies from $300$ meV to $200$~meV, while fixed at $300$ meV for the IPAFM and MTI blocks. It is evident magnetism in proximitized MTI stabilizes $M_z$ at the emergent interface between IPAFM and TI.}
\end{figure}

\noindent\textbf{Quantum Magnetism Simulation}

To better understand the proximity spin-orbit effects driving the complex magnetism in MBAT-BAT/MnTe, we conducted quantum magnetism simulations \cite{deCoster2021}, using realistic tight-binding and exchange parameters to model tri-material configurations consisting of generic IP ordered AFM (IPAFM), TI, and OOP aligned MTI blocks (see section III in SI for model details). As schematically shown in Fig.~\ref{fig:fig3}a (and Fig. S8), to match the experimental condition with $p$-doped chemical potential $\mu$, we have systematically varied the relative band alignment $E_0$ of the TI block, while for simplicity pinning that of the MTI block at $300$~meV and assuming a typical spin exchange interaction $J_{\rm ex}$ = 0.4 (0.04) eV in the IPAFM (MTI) block.

The ground-state energy of the entire structure is minimized allowing for the relaxation of the magnetization energy by self-consistently calculating the total energy of the system using direct diagonalization of the Green's function \cite{deCoster2021}. The simulations uncover that, under the modulation of a MTI block, significant OOP interlayer magnetic moments emerge at the IPAFM/TI interface, which is responsible for the magnetic switching at temperatures extending to $T_{\rm N}$ of IPAFM. Here, the topological surface states (TSS) sensitively depend on the band alignment (see Fig.~\ref{fig:fig3}b-d). Importantly for the AF and paramagnetic (PM) cases of the MTI block, a robust OOP magnetization and surface state gap exists even when the chemical potential is well below the Dirac point (Note: $E_0 = 200$ meV corresponds to the top of the valence bands), showing that even bulk conducting TIs will give rise to AHE when in the appropriate heterostructure configuration. It is of particular interest that there are nontrivial OOP magnetic moments at the IPAFM/TI interface even when the MTI block is in the PM (e.g. above $T_{\rm N}$ for MBAT) state. Such emergent OOP magnetization at atomically clean magnetic interfaces \cite{Katmis2016} enables novel magnetoelectric control for low power applications. 

We note that the actual structure deviates from our simplified simulation as an intercoupled natural heterostructure (see Fig.~\ref{fig:fig1}a). This simplified simulation is not intended to capture the exact stacking and transport properties of the measured device, but rather to describe the interfacial magnetism that can arise in such magnetic topological structures. Overall, the generic model sandwiching TI with IPAFM and MTI displaying distinct magnetic anisotropies offers crucial insights corroborating our magneto-transport and  quantitative two-component hysteresis model fitting \cite{Chen2020}.

\noindent\textbf{Spin-Orbit Torque Magnetic Switching} 

\begin{figure}
\includegraphics{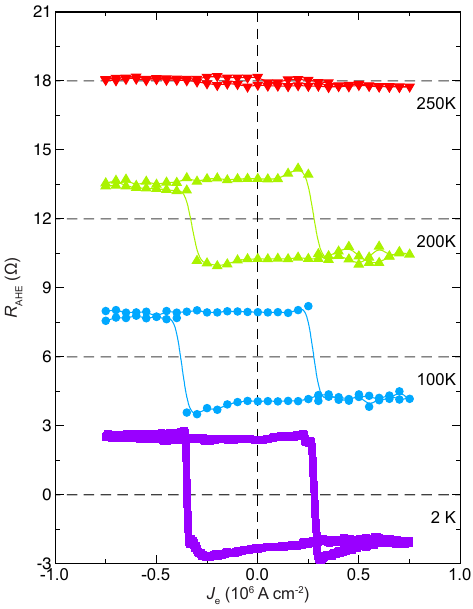}
\mycaption{\label{fig:fig4}\textbf{Spin-orbit torque magnetic switching.} External-magnetic-field-free SOT switching of the anomalous Hall resistance $R_{\rm AHE}$, driven by millisecond dc electrical current pulses at selected temperatures up to 250 K. Curves are shifted vertically by 6 $\Omega$ for clarity.}
\end{figure}

To explore nonvolatile magnetic switching in the MBAT-BAT/MnTe system, 1-ms writing current pulses $J_{\rm e}$ (with 1\% duty cycle to minimize thermal effects) were applied to provide the SOT, while monitoring the $R_{\rm AHE}$ by a standard lock-in technique using a 10 $\mu$A 17 Hz sine wave. Figure~\ref{fig:fig4} displays the typical switching profile at $T$ = 2 K, manifesting a low switching current density $J_{\rm c} = 3 \times 10^5$ Acm$^{-2}$ \cite{Han2017, Wu2019, Ye2022} with a typical channel thickness of 10 nm  and a resistivity of 2 m$\Omega$ cm. It is important to emphasize that the observed switching is independent of applied IP $H_x$ up to 16 T at 2 K (see Fig. S9), suggesting strong internal AF exchange bias fields favorably facilitate the deterministic switching. Furthermore, SOT switching is also demonstrated well above $T_{\rm N}$ for MBAT at 100 K and 200 K with little differences from the behavior at 2 K, which eventually diminishes at 250~K when approaching $T_{\rm N}$ of MnTe. 

The origin of the SOT is most reasonably attributed to the bulk spin Hall effect as the large carrier concentration in our sample indicates a chemical potential buried in the valence bands (see Fig. \ref{fig:fig3}). TSS coexist with bulk states at non-zero chemical potential, and therefore the topological Edelstein effect could also contribute to flipping an interfacial magnetization \cite{Fan2014}. However, given the large bulk carrier concentration, it is unlikely TSS play a large role in the observed switching.

The AHE shows a substantially reduced strength at 200 K (and no hysteresis), but the interfacial switching remains strong. These results are consistent with an interfacial magnetic moment that is stabilized by the interface up to the ordering temperature of the bulk MnTe moment. It appears that the ordering of the interfacial moment is the last to go upon rising temperature. This also accounts for the difference in AHE signal between magnetic flipping and SOT flipping: only the interface moment is affected by the SOT. The resulting external-magnetic-field-free deterministic SOT switching persists to $T_{\rm N}$ of a high-temperature AFM via proximity-modulated exchange coupling.
The platform leverages rich magnetic orders arising from coexisting IP and OOP moments coupled to a TI. 

\noindent\textbf{Conclusion}

The simultaneous interaction of multiple magnetic orders in spin-orbit coupled interfaces allows for non-volatile low power switching near room temperature in the absence of external magnetic field. The competing orders lead to an emergent platform, enabling exciting interface physics and new SOT applications. The rational incorporation of MBAT-based MTI with antiferromagnetically coupled BAT/MnTe interfaces opens opportunities for spin-orbit tunable topological spintronics. The BAT/MnTe heterostructure results in an AHE signature that has been investigated intently in its own right, but can be enhanced and deterministically torqued through proximity with MTI layers. While this natural heterostructure includes bulk-conducting BAT and stochastically formed MBAT, the resulting transport behaviors indicate that low current density, deterministic switching devices can be achieved if the MBAT-BAT/MnTe approach is optimized.

\section*{Methods}

\textbf{Sample growth.} \ch{Mn(Bi,Sb)2Te4}--\ch{(Bi,Sb)2Te3}/MnTe natural heterostructures with Bi$_2$Te$_3$ or \ch{Cr2Te3} buffer layers were grown in a MBE system under an ultrahigh-vacuum (UHV) environment of $10^{-10} - 10^{-9}$~Torr. Insulating \ch{Al2O3}(0001) and/or SrTiO$_3$(111) were used as substrates. The surface of \ch{Al2O3}(0001) was optimized by \emph{ex situ} chemical and thermal cleaning and \emph{in situ} outgassing at 800~$^\circ$C for 30~min.\ When using SrTiO$_3$(111), the passivated surface with atomic flatness was achieved by first annealing at 930~$^\circ$C for 3~hours in a tube furnace with flowing oxygen gas and then \emph{in situ} outgassed at 580~$^\circ$C for 30~min. After the initial surface preparation, the substrate temperature was lowered to 230~$^\circ$C for film growth, allowing enough surface mobility for the epitaxial crystallization of MnTe and \ch{(Bi,Sb)2Te3}. High purity (5N) source materials were co-evaporated from Knudsen cell evaporators (Bi, Sb and Te) and/or e-guns (Cr and Mn) as needed. The flux of each element was monitored by individual quartz crystals during the growth. Deposition of the 10--20 nm MnTe was conducted under a Te-rich condition with a typical Te/Mn flux ratio of 10, facilitated by a thin seeding layer of either 1 nm \ch{Cr2Te3} or 3 nm Bi$_2$Te$_3$ for improved crystallization. The \ch{(Bi,Sb)2Te3} layers were grown under the same Te-rich condition with Te/(Bi,Sb) flux ratio of 10 and Bi/Sb ratio for optimal chemical potential. Subsequently, 2 nm Te and/or 10 nm AlO$_x$ were capped \emph{in situ} at room temperature to protect the films from degradation by air exposure.  

\textbf{Structural characterizations.} X-ray diffraction (XRD) patterns (not shown) were obtained using a parallel beam of Cu K$_{\alpha1}$ radiation with wavelength $\lambda$ = 0.15406~nm in a Rigaku SmartLab system to confirm the structure and orientation. The $2\theta$ scan were typically between 10$^\circ$ and 120$^\circ$ with a step size of 0.05$^\circ$. X-ray reflectivity (XRR) measurements were performed at the Center for Nanophase Materials Sciences (CNMS), Oak Ridge National Laboratory, on a PANalytical X'Pert Pro MRD equipped with hybrid monochromator and Xe proportional counter. For XRR experiments, the X-ray beam was generated at 45~kV/40~mA, and the X-ray beam wavelength after the hybrid mirror was $\lambda$ = 0.15406~nm (Cu K$_{\alpha1}$ radiation). For microstructure analysis using cross-sectional electron microscopy, the specimens were prepared using a Thermo Scientific Helios G4 UX Dual Beam focused ion beam (FIB) system and examined in a JEOL ARM 200F STEM.

\textbf{Polarized neutron reflectometry.} PNR experiments were performed on the Magnetism Reflectometer at the Spallation Neutron Source at Oak Ridge National Laboratory, using neutrons with wavelengths $\lambda$ in a band of 0.2 -- 0.8~nm and a high polarization of 98.5--99{\%}. Measurements were carried out in a closed cycle refrigerator equipped with a 5~T superconducting magnet. Using the time-of-flight method \cite{pnrValeria, pnrSyromyatnikov2014JPCS_New, pnrJiang}, a collimated polychromatic beam of polarized neutrons with the wavelength band $\delta\lambda$ impinged on the film at a grazing incidence angle $\theta$, where it interacted with atomic nuclei and the spins of unpaired electrons. The reflected intensity $R^+$ and $R^-$ were measured as a function of momentum transfer, $Q = 4\pi\sin\theta/\lambda$, with the neutron spin parallel ($+$) or antiparallel ($-$), respectively, to the direction of the applied magnetic field. To separate the magnetic from the nuclear scattering, the spin asymmetry ratio SA = $(R^{+}(Q)-R^{-}(Q))/(R^{+}(Q) + R^{-}(Q))$ was obtained, with SA = 0 designating the absence of magnetic moment in the system. The spin-polarized neutrons, being electrically neutral, penetrate the entire heterostructure and probe simultaneously and nondestructively the structural, chemical and magnetic depth profiles of the multilayers and buried interfaces down to the substrate with a resolution of 0.5 nm \cite{pnrLauter-Pasyuk2007CS_Neutron, pnrLauter2012PSACR_215}. 

PNR is a 3D vector magnetometry technique that measures the direction and the absolute value of the magnetic moment and the lateral dimensions of magnetic domains. PNR scattering is used to directly measure the OOP component of magnetic moments as well \cite{Bai2021APL_Polarized,Zhu2012APL_Study,Zhu2016JPCS_Study,Kirichuk2025_Origin,Ji2011PRB_Perpendicular}. Measuring PMA using PNR involves analyzing the depth-dependent magnetization, specifically by comparing the sample's behavior under low IP fields (where moments are OOP) and high IP fields (where moments are forced IP). Because PNR is natively sensitive only to the IP component of magnetization, PMA is detected when the spin-up ($R^+$) and spin-down ($R^-$) reflectivities overlap at low fields, indicating that the magnetization is perpendicular to the film plane and does not interact with the IP neutron spin. 
Measurement in a very small magnetic field allows us to measure any possible component of the magnetization vector in plane of the sample or to prove the absence of such a component. Then if magnetization is OOP (PMA), neutrons polarized parallel or anti-parallel to the magnetic field experience the same magnetic potential, leading to no spin-flip scattering in the specular beam. If there is an IP component of magnetization, it will result in spin-flip scattering. If spin-flip scattering is present it contributes to the reflectivity \cite{Blundell1995PRB_Spinorientation,Fields2026AAMI_NonAltermagnetic} and/or off-specular scattering \cite{Toperverg2001PBCM_3D, pnrLauter2012PSACR_215}.

\textbf{Device fabrication and transport measurement.} Samples were fabricated using electron beam lithography, Argon plasma etching, and Ti/Au metallization. Magnetotransport experiments were performed in an 18 Tesla Cryogen Free Measurement System from Cryogenics Ltd.

\textbf{Data availability.} The data that support the findings of this study are available from the corresponding authors upon reasonable request.

\section*{Acknowledgments}
H.C. acknowledges the support of the Army Research Office (ARO, W911NF-25-1-0215), the Canada Research Chairs (CRC) Program and the Natural Sciences and Engineering Research Council of Canada (NSERC), Discovery Grant RGPIN-2024-06497 and ALLRP 592642-2024. The work at MIT was supported by the ARO (W911NF-20-2-0061 and DURIP W911NF-20-1-0074), National Science Foundation (NSF-DMR 1700137) and Office of Naval Research (N00014-20-1-2306). J.S.M., S.-W.W. and Y.H. thank the Center for Integrated Quantum Materials (NSF-DMR 1231319) for financial support. S.-W.W. acknowledges the support from the National Science and Technology Council (Grant No. NSTC 112-2112-M-006-038-MY3) in Taiwan. D.Hei. thanks support from NSF grant DMR-1905662 and the Air Force Office of Scientific Research award FA9550-20-1-0247. T.B.E. was partially supported by NSF under Grant No. DGE 1633587. The electron microscopy was performed at the Analytical Instrumentation Facility (AIF) at North Carolina State University, which is supported by the State of North Carolina and NSF (Award No. ECCS-2025064). The AIF is a member of the North Carolina Research Triangle Nanotechnology Network (RTNN), a site in the National Nanotechnology Coordinated Infrastructure (NNCI). This work made use of the MIT Materials Research Laboratory. This research used resources at the Spallation Neutron Source, a Department of Energy Office of Science User Facility operated by the Oak Ridge National Laboratory (ORNL). Neutron reflectometry measurements, beamtime proposal IPTS-28390, were carried out on the Magnetism Reflectometer at the SNS, which is sponsored by the Scientific User Facilities Division, Office of Basic Energy Sciences, DOE. XRR measurements were conducted at the Center for Nanophase Materials Sciences (CNMS) at ORNL, which is a DOE Office of Science User Facility. The United States Government retains, and the publisher, by accepting the article for publication, acknowledges that the United States Government retains, a nonexclusive, paid-up, irrevocable, worldwide license to publish or reproduce the published form of this manuscript, or allow others to do so, for United States Government purposes. The Department of Energy will provide public access to these results of federally sponsored research in accordance with the DOE Public Access Plan (http://energy.gov/downloads/doe-public-access-plan).

\section*{Author contributions}
H.C., O.A.V., V.L. and J.S.M. conceived the research. H.C. grew the samples, measured the magnetic properties with D.Hei, and analyzed the structure with J.-F.M.. T.B.E., W.G. and W.L.S. performed STEM imaging. O.A.V., S.-W.W., Y.H., D.Het. and H.C. carried out device fabrication, transport measurement and analysis. J.K. performed XRR measurements, V.L. and H.A. conducted PNR experiments, V.L. analyzed XRR and PNR data. G.J.d.C. and M.J.G. provided theoretic input. O.A.V. and H.C. wrote the paper with input from all authors, who discussed the results.


\bibliography{0-MS}
\bibliographystyle{achemso}

\end{document}